\documentclass[journal]{IEEEtran}

\usepackage{graphicx}
\usepackage{amssymb,amsmath}
\usepackage{multirow}
\usepackage{tabularx}
\usepackage{slashbox}
\usepackage{color}
\usepackage{lettrine}
\usepackage{mdwlist}
\usepackage{cite}

\begin{document}
	\title{SPAD-Based Optical Wireless Communication with ACO-OFDM}
	\author{Shenjie Huang, Cheng Chen, Mohammad Dehghani Soltani, Robert Henderson, Harald Haas, and Majid Safari 
	\thanks{This work was supported by Engineering and Physical Sciences Research Council (EPSRC) under Grant EP/S016570/1 (TOWS). (Corresponding author: Shenjie Huang.)}
	\thanks{S. Huang, M. D. Soltani, R. Henderson, and M. Safari are with the School of Engineering, the University of Edinburgh, Edinburgh EH9 3JL, UK. C. Chen and H. Haas are with LiFi Research and Development Centre, University of Strathclyde, Glasgow G1 1RD, UK. (e-mail: shenjie.huang@ed.ac.uk; c.chen@strath.ac.uk; m.dehghani@ed.ac.uk; robert.henderson@ed.ac.uk; harald.haas@strath.ac.uk; majid.safari@ed.ac.uk).}
	\vspace{-0.5cm}
	}

	\maketitle
	\begin{abstract}      
	
	The sensitivity of the optical wireless communication (OWC) can be effectively improved by employing the highly sensitive single-photon avalanche diode (SPAD) arrays. However, the nonlinear distortion introduced by the dead time strongly limits the throughput of the SPAD-based OWC systems. Optical orthogonal frequency division multiplexing (OFDM) can be employed in the systems with SPAD arrays to improve the spectral efficiency. In this work, a theoretical performance analysis of SPAD-based OWC system with asymmetrically-clipped optical OFDM (ACO-OFDM) is presented. 
%	The analytical expressions of both signal-to-noise ratio (SNR) and bit error rate (BER) are derived and validated. 
	The impact of the SPAD nonlinearity on the system performance is investigated. In addition, the comparison of the considered scheme with direct-current-biased optical OFDM (DCO-OFDM) is presented showing the distinct reliable operation regimes of the two schemes. In the low power regimes, ACO-OFDM outperforms DCO-OFDM; whereas, the latter is more preferable in the high power regimes. 
	\end{abstract}
\vspace{-0.1cm}
	\begin{IEEEkeywords}
		Optical wireless communication, orthogonal frequency division multiplexing, single-photon avalanche diode. 
	\end{IEEEkeywords}
\vspace{-0.2cm}
\section{Introduction}

%It is forecasted that by 2023 there will be 5.7 billion connected mobile users \cite{cisco}, which will makes the conventional radio frequency (RF) wireless network face significant challenges such as spectrum shortage and access congestion. 

In recent decades, the optical wireless communication (OWC) has been continuously gaining interest in both industry and academia and is considered as a potential candidates to provide more powerful wireless connections in the future. The performance of OWC can be strongly degraded by the occasional outages introduced by multiple effects such as adverse weather condition and user mobility in outdoor and indoor scenarios, respectively. One effective way of improving the performance of OWC systems under weak power reception is employing highly sensitive photon counting receivers such as a single-photon avalanche diode (SPAD). A SPAD receiver is achieved by biasing a traditional linear photodiode above the breakdown voltage so that it operates in the `Geiger mode' \cite{Huang22}. When a photon is received by SPAD receivers, an avalanche is triggered generating a striking electrical output pulse which realizes the single photon detection.

Although a SPAD receiver has photon count capability, after each avalanche
it has to be quenched for a short period of time when it becomes blind to
any incident photon arrivals, which is also know as the \textit{dead time}. The throughputs of the OWC systems with SPAD receivers are strongly limited by the nonlinearity induced by the dead time \cite{Huang22}. Although most of the prior SPAD-based OWC works focus on the on-off keying (OOK) \cite{Matthews21}, some works have been conducted to investigate the application of optical orthogonal frequency division multiplexing (OFDM) in SPAD-based OWC systems to improve the spectral efficiency \cite{Huang:22,Zhang22}. In particular, in \cite{Huang:22} a record data rate of $5$ Gbps is achieved experimentally using a commercial SPAD receiver with the employment of the optical OFDM and nonlinear equalizer. Despite the aforementioned experimental works, a theoretical performance analysis of SPAD-based OWC systems with direct-current-biased optical OFDM (DCO-OFDM) was conducted very recently \cite{arxivHuang}. Besides DCO-OFDM, asymmetrically-clipped optical OFDM (ACO-OFDM) is another commonly used optical OFDM scheme which does not require a DC bias and enjoys a better power efficiency \cite{Armstrong08}. It is concluded that ACO-OFDM is well suited to some practical applications such as the visible light communication (VLC) with dimming control.  However, to the best of our knowledge, a complete performance analysis of SPAD-based OWC systems with ACO-OFDM is still missing. In this work, we aim to fill this research gap. The analytical expressions of the signal-to-noise ratio (SNR) and bit error rate (BER) of the considered system are derived. The influence of the unique SPAD nonlinearity on the system performance is investigated. In addition, an in-depth comparison with SPAD DCO-OFDM is also presented. 

\section{SPAD-Based ACO-OFDM System}\label{OFDMsys}
%\begin{figure*}[!t]
%	\begin{center}
%		\includegraphics[width=0.8\textwidth]{system_ACO.eps}
%		\caption{Block diagram of the SPAD-based OWC with OFDM.}\label{system_figure}
%	\end{center}
%\end{figure*}
\subsection{ACO-OFDM Transmission}
%Fig. \ref{system_figure} presents the block diagram of a SPAD-based OWC system with ACO-OFDM. 

For a SPAD-based OWC system with ACO-OFDM, at the transmitter, the input bit stream is transformed into a complex symbol stream by the $M$-quadrature amplitude modulation (QAM) modulator, where $M$ denotes the constellation size. The symbol stream is then serial-to-parallel (S/P) converted to form vectors suitable for inverse fast Fourier transform (IFFT) operation. Considering a $K$-point fast Fourier transform (FFT) operation, only the odd subcarriers of the first half of the OFDM frame with index $k=1, 3, 5, \cdots, K/2-1$ are used to carry the information, whereas the even subcarriers are left unused. Therefore, the number of information carrying subcarriers is $K'=K/4$. Hermitian symmetry is applied to the rest of the OFDM frame in order to obtain the real-valued symbols  after the  IFFT operation. Denote the generated OFDM frame as $X[k]$, the time-domain signal $x[n]$ can be obtained after the IFFT as $x[n]=\frac{1}{\sqrt{K}} \left(\sum_{k=0}^{K-1} X[k] e^{\frac{2\pi nkj}{K}}\right)$. According to the central limit theorem (CLT), the amplitude of $x[n]$ is approximately zero-mean Gaussian distributed when $K$ is relatively large \cite{Tsonev13}. Considering the uniform power allocation over the subcarriers, the variance of signal should be $\sigma_X^2=2$ to ensure that $x[n]$ is with unit variance \cite{Dimitrov12}.

Since only the odd subcarriers are utilized, the time-domain signal $x[n]$ has the following anti-symmetry
\begin{equation}\label{xn}
	x[n]=-x\left[n+{K}/{2}\right]\;\; \mathrm{for} \;\;\; n \in \left[0,{K}/{2}-1\right].
\end{equation} 
Therefore, the clipping of all negative samples at the transmitter does not introduce any information loss and the information can still be successfully decoded at the receiver \cite{Dimitrov12}. Considering that the peak-to-average power ratio (PAPR) of the generated signal $x[n]$ is relatively high whereas practical light sources are with limited dynamic ranges, $x[n]$ should also be properly clipped at a top clipping level $\kappa$. Hence, the clipped signal can be expressed as 
\begin{equation}\label{xclipped}
	x_{\rm c}[n]=
	\begin{cases}
		\kappa, & \mathrm{if}\qquad  x[n]\geq \kappa,\\
		x[n],     & \mathrm{if}\qquad  0<x[n]< \kappa,\\
		0, & \mathrm{if}\qquad  x[n]\leq 0. 
	\end{cases}
\end{equation}
After applying scaling and digital-to-analog conversion, the resultant electrical signal is used to drive the light source. In effect, the optical power of the $n$th time-domain OFDM sample emitted from the source is given by 
$x_{\rm t}[n]=\xi\, x_{\rm c}[n]$ where $\xi$ denotes the scaling factor. Denoting the maximal optical power of the light source as $P_{\rm max}$, $\xi \kappa=P_{\rm max}$ should be satisfied, which leads to $\xi={P_{\rm max}}/{\kappa}$.
The average transmit optical power is given by 
\begin{equation}\label{PTx}
	\overline{P}_{\rm Tx}=\xi\left[{1}/{\sqrt{2\pi}}   -f_N(\kappa)+\kappa\, Q(\kappa)\right],
\end{equation}
where 
%\begin{equation}\label{fx}
%f(x)=\frac{1}{\sqrt{2\pi}}\exp\left(-\frac{x^2}{2}\right),
%\end{equation}
$f_N(x)$ is the  probability density function (PDF) of a standard Gaussian distribution and ${Q}(\cdot)$ denotes the Q-function. 
\vspace{-0.2cm}
\subsection{SPAD Receivers}
%A SPAD operates in the `Geiger' mode and can generate a very large current by receiving a single photon making it a photon counting receiver. 
The photodetection process of an ideal photon counter can be modelled using Poisson statistics. However, the performance of the practical SPAD-based receivers suffer from the non-ideal effects such as dead time, photon detection efficiency (PDE), dark count rate (DCR), afterpulsing and crosstalk. To mitigate the significant nonlinearity effects introduced by the dead time and improve the photon counting capability, arrays of SPADs are commonly used in OWC \cite{Matthews21,Huang22}. Considering that the channel loss is $\zeta$, the average received signal optical power is given by $\overline{P}_{\rm Rx}=\zeta\,\overline{P}_{\rm Tx}$. Assuming a precise time synchronization between transceivers, the received optical power when the $n$th OFDM sample is transmitted can be expressed as $P_{\rm Rx}[n]=\zeta x_{\rm t}[n]$. The corresponding incident photon rate of the SPAD array  is \cite{Khalighi20}
\begin{equation}\label{lama2}
	\lambda_\mathrm{a}[n]= \mathcal{C}_{\rm s}x_{\rm t}[n]+\mathcal{C}_{\rm n}.
\end{equation}
where 
\begin{equation}
	\begin{cases}
		\mathcal{C}_{\rm s}=\Upsilon_{\rm PDE}\zeta\left(1+\varphi_{\rm AP}+\varphi_{\rm CT}\right)/E_{\rm ph},\\[2pt]
		\mathcal{C}_{\rm n}=(\vartheta_{\rm DCR}+\vartheta_{\rm B})\left(1+\varphi_{\rm AP}+\varphi_{\rm CT}\right),
	\end{cases}
\end{equation}
$\Upsilon_{\rm PDE}$ is the PDE of the SPAD, $E_{\rm ph}$ is the photon energy, $\vartheta_{\rm B}$ denotes the background photon rate, and $\vartheta_{\rm DCR}$, $\varphi_{\rm AP}$ and $\varphi_{\rm CT}$ refer to the DCR of the array, the probabilities of afterpulsing and crosstalk, respectively. The photon rate $\vartheta_{\rm B}$ equals to $\Upsilon_{\rm PDE}P_{\rm B}/E_{\rm ph}$ where $P_{\rm B}$ is the ambient light power.

There are two main types of SPADs, i.e., active quenched (AQ) and passive quenched (PQ) SPADs. The latter benefit from the simpler circuit design and higher PDE making them commonly employed in the commercial products \cite{Matthews21}. In this work, we hence consider that the employed SPAD receiver is PQ-based. At the receiver, the SPAD array detector outputs the detected photon count during every OFDM sample duration $T_{\rm s}$. 
Denote that, after the S/P mapping, the detected photon count of the SPAD array when the $n$th OFDM sample is transmitted as $y[n]$. When the array size is relatively large, according to the CLT, $y[n]$ is approximately Gaussian distributed with mean and variance given by \cite{omote,Huang22}  
\begin{equation}\label{meanPQ}
	\mu_{\rm a}(x[n])={\lambda_\mathrm{a}[n]T_{\rm s}} \exp\left(-\frac{\lambda_\mathrm{a}[n]\tau_{\rm d}}{N_{\rm a}}\right),
\end{equation}	
and 
\begin{align}\label{varPQ}
	\sigma_{\rm a}^2 (x[n])=&\lambda_\mathrm{a}[n]T_{\rm s} \exp\left(-\frac{\lambda_\mathrm{a} [n]\tau_{\rm d}}{N_{\rm a}}\right)\\
	&-\frac{\lambda_\mathrm{a}^2 [n]T_{\rm s}\tau_{\rm d}}{N_{\rm a}} \exp\left(-\frac{2\lambda_\mathrm{a} [n]\tau_{\rm d}}{N_{\rm a}}\right)\left(2-\frac{\tau_{\rm d}}{T_{\rm s}}\right),\nonumber
\end{align}
respectively, where $N_{\rm a}$ denotes the number of SPADs in the array and $\tau_{\rm d}$ refers to the dead time. Note that (\ref{meanPQ}) indicates that 
with the increase of the incident photon rate $\lambda_\mathrm{a}[n]$, the detected photon count firstly increases and then decreases, hence 
the received optical signal is nonlinearly distorted by the SPAD receiver. The detected photon count $y[n]$ can be written by
\begin{equation}\label{yn1}
	y[n]=\mu_{\rm a}(x[n])+w_{\rm s}[n],
\end{equation} 
where $w_{\rm s}[n]$ represents the shot noise which is Gaussian distributed with zero mean and signal dependent variance given by (\ref{varPQ}). 
%As shown in Fig. \ref{system_figure}, 
At the receiver, the signal $y[n]$ is then converted back to the frequency-domain using the FFT operation given by $Y[k]=\frac{1}{\sqrt{K}} \left(\sum_{n=0}^{K-1} y[n] e^{-\frac{2\pi nkj}{K}}\right)$.
Finally, after the single-tap equalization, P/S mapping, and QAM demodulation, the recovered bit steam can be achieved. 
\vspace{-0.1cm}
\section{Theoretical Analysis of SPAD ACO-OFDM}\label{theore} 
In the considered SPAD OFDM system, two nonlinear distortions exist. The first is the clipping-induced distortion as presented in (\ref{xclipped}), which also exists in standard OFDM-based OWC systems with linear receivers \cite{Dimitrov12}.  The second is the additional unique SPAD-induced distortion given in (\ref{meanPQ}).  We combine these two nonlinear distortions and investigate the system performance in the presence of the effective nonlinear distortion. By substituting (\ref{xclipped}) and (\ref{lama2}) into (\ref{meanPQ}), the combined nonlinear distortion of the transmitted signal $x[n]$ is given by
\begin{align}\label{distortPQ}
	&\mu_{\rm a}(x[n])=\\
	&\begin{cases}
		\left(\psi_1\kappa+\mathcal{C}_{\rm n}\right)	T_{\rm s} \exp\left[-\frac{\left(\psi_1\kappa+\mathcal{C}_{\rm n}\right)	\tau_{\rm d}}{N_{\rm a}}\right], & \! \mathrm{if}\,  x[n]\geq \kappa,\\[1.5ex]
		\left(\psi_1x[n]+\mathcal{C}_{\rm n}\right)T_{\rm s} \exp\left[-\frac{\left(\psi_1x[n]+\mathcal{C}_{\rm n}\right)	\tau_{\rm d}}{N_{\rm a}}\right],  & \! \mathrm{if}\,  0<x[n]< \kappa,\\[1.5ex]
		{\mathcal{C}_{\rm n}	T_{\rm s}} \exp\left(-\frac{\mathcal{C}_{\rm n}	\tau_{\rm d}}{N_{\rm a}}\right), & \! \mathrm{if}\,  x[n]\leq 0. 
	\end{cases}\nonumber
\end{align}
where $\psi_1=\mathcal{C}_{\rm s}{P_{\rm max}}/{\kappa}$.

According to the Bussgang theorem, the nonlinear distortion in an OFDM-based system can be described by a gain factor ($\alpha$) and an additional signal-independent distortion-induced noise ($w_{\rm d}[n]$) \cite{Tsonev13,Dimitrov12} as
\begin{equation}\label{BT1}
	\mu_{\rm a}(x[n])= \alpha x[n] + w_{\rm d}[n].
\end{equation}
The gain factor $\alpha$ can be calculated as \cite{arxivHuang}
\begin{align}\label{alpha_fin}
	\alpha=&\frac{\psi_1^2\tau_{\rm d}T_{\rm s}}{\sqrt{2\pi}N_{\rm a}}\left\{\exp \left[-\frac{\kappa^2}{2}-\frac{\tau_{\rm d}}{N_{\rm a}}\left(\psi_1\kappa+\mathcal{C}_{\rm n}\right)\right]
	-e^{-\frac{\tau_{\rm d}\mathcal{C}_{\rm n}}{N_{\rm a}}}\right\}\nonumber\\[0.5em]
	&\;\;+{\psi_1}\,T_{\rm s}\,e^{-\frac{\mathcal{C}_{\rm n}\tau_{\rm d}}{N_{\rm a}}+\frac{\psi_{1}^2\tau_{\rm d}^2}{2N_{\rm a}^2}}\left[1+\frac{\psi_{1}^2\tau_{\rm d}^2}{N_{\rm a}^2}-\frac{\mathcal{C}_{\rm n}\tau_{\rm d}}{N_{\rm a}}\right]\nonumber\\
	&\;\;\times \left[Q\left(\frac{\psi_{1}\tau_{\rm d}}{N_{\rm a}}\right)-Q\left(\kappa+\frac{\psi_{1}\tau_{\rm d}}{N_{\rm a}}\right)\right].
\end{align} 
The variance of $w_{\rm d}[n]$, denoted as $\sigma_{w_{\rm d}}^2$, is given by
\begin{equation}\label{sigma_wd}
	\sigma_{w_{\rm d}}^2=\mathbb{E}\left\{\mu^2_{\rm a}(x[n])\right\}-\mathbb{E}^2\left\{\mu_{\rm a}(x[n])\right\}-\alpha^2,
\end{equation}
where $\mathbb{E}\{\cdot\}$ denotes the statistical expectation. The two moments of $\mu_{\rm a}(x[n])$ in (\ref{sigma_wd}) can be found in  \cite{arxivHuang}. 

By plugging (\ref{BT1}) into (\ref{yn1}), the SPAD output $y[n]$ can be rewritten as
\begin{equation}\label{yn2}
	y[n]=\alpha x[n]+w_{\rm d}[n]+w_{\rm s}[n].
\end{equation}
After applying FFT operation, the signal in the frequency domain can be expressed as
\begin{equation}\label{yn3}
	Y[k]=\alpha X[k]+W_{\rm d}[k]+W_{\rm s}[k],
\end{equation}
where $W_{\rm d}[k]$ and $W_{\rm s}[k]$ denote the FFT of $w_{\rm d}[n]$ and $w_{\rm s}[n]$, respectively. When the number of subcarriers is sufficiently large, CLT applies and both $W_{\rm d}[k]$ and $W_{\rm s}[k]$ are zero-mean Gaussian noise terms \cite{Tsonev13}.  
As a result, in the frequency domain the received signal is the transmitted signal multiplied by a gain factor plus two additive Gaussian noises.  The variance of the shot noise  $W_{\rm s}[k]$ for the considered system with ACO-OFDM, denoted as $\sigma_{W_{\rm s}}^2$, is identical to that with DCO-OFDM which has been derived in our previous work \cite{arxivHuang}. Now let's derive the variance of the distortion-induced noise in the frequency domain $W_{\rm d}[k]$. 
The variance of $W_{\rm d}[k]$  can be expressed as (\ref{sigmaWd}) on the top of next page 
\begin{figure*}[!t]
\begin{align}\label{sigmaWd}
	\sigma_{W_{\rm d}}^2[k]&=\mathbb{E}\{|W_{\rm d}[k]|^2\}=\frac{1}{K}\sum_{n=0}^{K-1}\sum_{m=0}^{K-1}\mathbb{E}\{ w_{\rm d}[n]w_{\rm d}[m]\}e^{-\frac{2\pi nkj}{K}+\frac{2\pi mkj}{K}}\nonumber \\[-2pt]
	&=\frac{1}{K}\sum_{n=0}^{K-1}\mathbb{E}\{w^2_{\rm d}[n]\}+\frac{1}{K}\sum_{n=0}^{K-1}\mathbb{E}\{w_{\rm d}[n]w_{\rm d}[n_{\rm f}]\}e^{-\frac{2\pi nkj}{K}+\frac{2\pi n_{\rm f}kj}{K}}-\frac{1}{K}\sum_{n=0}^{K-1}\mathbb{E}^2\{w_{\rm d}[n]\}\,e^{-\frac{2\pi nkj}{K}}\sum_{{m\neq n, n_{\rm f} }}^{K-1}e^{\frac{2\pi mkj}{K}},\nonumber\\[-2pt]
	&=\sigma_{w_{\rm d}}^2+\mathbb{E}^2\{w_{\rm d}[n]\}+\frac{\mathbb{E}\{w_{\rm d}[n]w_{\rm d}[n_{\rm f}]\}}{K}\underbrace{\sum_{n=0}^{K-1}e^{-\frac{2\pi nkj}{K}+\frac{2\pi n_{\rm f}kj}{K}}}_{\mathcal{T}_1}-\frac{\mathbb{E}^2\{w_{\rm d}[n]\}}{K}\underbrace{\sum_{n=0}^{K-1}\,e^{-\frac{2\pi nkj}{K}}\sum_{{m\neq n, n_{\rm f} }}^{K-1}e^{\frac{2\pi mkj}{K}}}_{\mathcal{T}_2}.
\end{align}
\vspace{-0.6cm}
\hrulefill
\end{figure*}
where $n_{\rm f}$ denotes the anti-symmetrical index of $n$ as 
\begin{equation}\label{xcfold}
	n_{\rm f}=
	\begin{cases}
		n+\frac{K}{2} &  \text{if} \quad n\leq  \frac{K}{2}-1,\\
		n-\frac{K}{2} &  \text{if} \quad n> \frac{K}{2}-1.
	\end{cases}		
\end{equation}
Note that due to the anti-symmetry property of the time-domain ACO-OFDM samples shown in (\ref{xn}), $w_{\rm d}[n]$ and $w_{\rm d}[n_{\rm f}]$ are correlated. In fact, because of the existence of this correlation, $\sigma_{W_{\rm d}}^2[k]$ does not simply equal to $\sigma_{w_{\rm d}}^2$, which differs from the system with DCO-OFDM \cite{arxivHuang}.  The term $\mathcal{T}_1$ in (\ref{sigmaWd}) can be calculated as $\mathcal{T}_1=\sum_{n=0}^{\frac{K}{2}-1}e^{\pi kj}+\sum_{n=\frac{K}{2}}^{K-1}e^{-\pi kj}=K \cos(\pi k)$.
%\begin{align}
%	\mathcal{T}_1=\sum_{n=0}^{\frac{K}{2}-1}e^{\pi kj}+\sum_{n=\frac{K}{2}}^{K-1}e^{-\pi kj}=K \cos(\pi k).
%\end{align}
On the other hand for $k\geq 1$ the term $\mathcal{T}_2$ in (\ref{sigmaWd}) can be calculated as 
\begin{align}
\!\!\!\!\mathcal{T}_2\!&=\!-\!\!\sum_{n=0}^{K-1}\!\!e^{-\frac{2\pi nkj}{K}}\!\!\left(e^{\frac{2\pi nkj}{K}}\!\!+\!e^{\frac{2\pi n_{\rm f}kj}{K}}\right)\!=\!-K\!-\!\!\!\sum_{n=0}^{K-1}\! e^{-\frac{2\pi (n-n_{\rm f})kj}{K}}\nonumber\\[-4pt]
	&=-K\!-\!\!\sum_{n=0}^{\frac{K}{2}-1} e^{\pi kj}-\!\!\sum_{n=\frac{K}{2}}^{K-1} e^{-\pi kj}=\!-K\!-\!K\cos(\pi k).
\end{align}
Both $\mathcal{T}_1$ and $\mathcal{T}_2$ are distinct for odd and even subcarriers, hence the distortion-induced noise variances are different for these subcarriers. For information carrying odd $k$, one can get $\mathcal{T}_1=-K$ and $\mathcal{T}_2=0$. Therefore, the variance of the distortion-induced noise in the frequency domain can be expressed as
\begin{equation}\label{sigmaWd2}
	\sigma_{W_{\rm d}}^2=\sigma_{w_{\rm d}}^2+\mathbb{E}^2\{w_{\rm d}[n]\}-\mathbb{E}\{w_{\rm d}[n]w_{\rm d}[n_{\rm f}]\}.
\end{equation}
Invoking the definition of $w_{\rm d}[n]$ in (\ref{BT1}), one has $\mathbb{E}\{w_{\rm d}[n]\}=\mathbb{E}\{\mu_{\rm a}(x[n])\}$. Substituting (\ref{sigma_wd}) into (\ref{sigmaWd2}) results in
\begin{equation}\label{sigmaWd3}
	\sigma_{W_{\rm d}}^2=\mathbb{E}\left\{\mu^2_{\rm a}(x[n])\right\}-\alpha^2-\mathbb{E}\{w_{\rm d}[n]w_{\rm d}[n_{\rm f}]\}.
\end{equation}
The moment $\mathbb{E}\left\{\mu^2_{\rm a}(x[n])\right\}$ can be found in \cite{arxivHuang}. For the correlation term $\mathbb{E}\{w_{\rm d}[n]w_{\rm d}[n_{\rm f}]\}$, one can get
\begin{align}\label{Ewdwdf}
	\mathbb{E}\{w_{\rm d}[n]&w_{\rm d}[n_{\rm f}]\}=\\[2pt]
	&\mathbb{E}\left\{\mu_{\rm a}(x[n]) \mu_{\rm a}(x[n_{\rm f}])\right\}-\alpha \mathbb{E}\left\{ \mu_{\rm a}(x[n])x[n_{\rm f}] \right\}\nonumber\\[2pt]
	&-\alpha \mathbb{E}\left\{ \mu_{\rm a}(x[n_{\rm f}])x[n] \right\}+\alpha^2\mathbb{E}\left\{ x[n] x[n_{\rm f}] \right\}.\nonumber
\end{align}
From (\ref{xn}) and (\ref{xcfold}), the equation $x[n_{\rm f}]=-x[n]$ holds and substituting this into (\ref{Ewdwdf}) leads to 
\begin{align}\label{Ewdwdf2}
	\mathbb{E}&\{w_{\rm d}[n]w_{\rm d}[n_{\rm f}]\}=\\[2pt]
	&\mathbb{E}\left\{\mu_{\rm a}(x) \mu_{\rm a}(-x)\right\}+\alpha \mathbb{E}\left\{ \mu_{\rm a}(x)x \right\}-\alpha \mathbb{E}\left\{\mu_{\rm a}(-x)x \right\}-\alpha^2,\nonumber
\end{align}
where the sample index $n$ is dropped for simplicity. Since the term $\mathbb{E}\left\{ \mu_{\rm a}(x)x \right\}$  and $\mathbb{E}\left\{\mu_{\rm a}(-x)x \right\}$  equals to $\alpha$ and $-\alpha$, respectively, expression (\ref{Ewdwdf2}) can be rewritten as 
\begin{equation}\label{Ewdwdf3}
	\mathbb{E}\{w_{\rm d}[n]w_{\rm d}[n_{\rm f}]\}=\mathbb{E}\left\{\mu_{\rm a}(x) \mu_{\rm a}(-x)\right\} +\alpha^2.
\end{equation}
After some mathematical manipulations, the term $\mathbb{E}\left\{\mu_{\rm a}(x) \mu_{\rm a}(-x)\right\}$ in (\ref{Ewdwdf3}) can be calculated as
\begin{align}\label{Euaua}
	\mathbb{E}&\{\mu_{\rm a}(x) \mu_{\rm a}(-x)\}=\\
	&2\mathcal{C}_{\rm n}T_{\rm s}^2e^{-\frac{(\psi_1\kappa+2\mathcal{C}_{\rm n})\tau_{\rm d}}{N_{\rm a}}}\left(\psi_1\kappa+\mathcal{C}_{\rm n}\right)Q\left(\kappa\right)+2\mathcal{C}_{\rm n}T_{\rm s}e^{-\frac{\mathcal{C}_{\rm n}\tau_{\rm d}}{N_{\rm a}}}\mathcal{T}_3,\nonumber
\end{align}
where $\mathcal{T}_3=\int_{0}^{\kappa}\mu_{\rm a}(x)f_N(x)\mathrm{d} x$, which can be solved analytically as
\begin{align}\label{calM}
	&\mathcal{T}_3=\frac{T_{\rm s}\psi_{1}}{\sqrt{2\pi}}\,e^{-\frac{\mathcal{C}_{\rm n}\tau_{\rm d}}{N_{\rm a}}}\left[1-e^{-\frac{1}{2}\kappa^2-\frac{\psi_{1}\tau_{\rm d}\kappa}{N_{\rm a}}  }\right]\\[0.4em]
	&\,+\!{T_{\rm s}}\!\left(\!\frac{\tau_{\rm d}\psi_1^2}{N_a}\!-\!\mathcal{C}_{\rm n}\!\right)\!e^{-\frac{\mathcal{C}_{\rm n}\tau_{\rm d}}{N_{\rm a}}+\frac{\psi_1^2\tau_{\rm d}^2}{2N_{\rm a}^2}}\!\!\left[\!{Q}\!\left(\!{\kappa\!+\!\frac{\psi_{1}\tau_{\rm d}}{N_{\rm a}}}\!\right)\!-\!{Q}\!\left(\!\frac{\psi_{1}\tau_{\rm d}}{N_{\rm a}}\!\right)\right].\nonumber
\end{align}
Finally, by substituting (\ref{Euaua}) and (\ref{Ewdwdf3}) into (\ref{sigmaWd3}), the analytical expression of $\sigma_{W_{\rm d}}^2$ is achieved. For the special case of ideal photon counting receiver ($\tau_{\rm d}=0$), the variance of the distortion-induced noise in the frequency domain can be simplified as 
 \begin{equation}
 	\sigma_{W_{\rm d, id}}^2=\psi_{1}^2T_{\rm s}^2\left[Q(\kappa)-2Q^2(\kappa)-\kappa f(\kappa)+\kappa^2Q(\kappa)\right],
 \end{equation}
which is in line with the derived clipping noise of ACO-OFDM systems with linear receivers \cite{Dimitrov12}. 

Since both noise terms in (\ref{yn3}) are uncorrelated with the signal making it a standard additive Gaussian noise channel model, the SNR of the received signal is given by
\begin{equation}\label{gamma}
	\gamma=\frac{\alpha^2\sigma_X^2}{\sigma_{W_{\rm d}}^2+\sigma_{W_{\rm s}}^2}=\frac{1}{\frac{1}{\gamma_{\rm d}}+\frac{1}{\gamma_{\rm s}}}.
\end{equation}
where we denote the terms $\gamma_{\rm d}={2\alpha^2}/{\sigma_{W_{\rm d}}^2}$ and $\gamma_{\rm s}={2\alpha^2}/{\sigma_{W_{\rm s}}^2}$
%\begin{equation}
%	\gamma_{\rm d}=\frac{2\alpha^2}{\sigma_{W_{\rm d}}^2}, \quad \mathrm{and} \quad \gamma_{\rm s}=\frac{2\alpha^2}{\sigma_{W_{\rm s}}^2},
%\end{equation}
the signal-to-distortion-noise ratio (SDNR) and signal-to-shot-noise ratio (SSNR), respectively, where $\alpha$ and $\sigma_{W_{\rm d}}^2$ are given by (\ref{alpha_fin}) and (\ref{sigmaWd3}), respectively. The derivation of the analytical expression of the SNR is now complete. 
%The term $2\alpha^2$ refers to the received electrical signal power, therefore the absolute value of the gain factor, i.e., $|\alpha|$, can be used to measure the signal power. 
Equation (\ref{gamma}) indicates that SNR depends on two factors, i.e., SDNR and SSNR. The BER of the considered system can be achieved by substituting $\gamma$ into the QAM BER equation, e.g., \cite[eq. (30)]{Dimitrov12}.

%\begin{align}\label{BER}
%	P_e=\frac{4\big(\sqrt{M}-1\big)}{\sqrt{M}\mathrm{log}_2(M)}&{Q} \Bigg(\sqrt{\frac{3\gamma}{M-1}}\Bigg)\\
%	&+\frac{4\big(\sqrt{M}-2\big)}{\sqrt{M}\mathrm{log}_2(M)}{Q} \Bigg(3\sqrt{\frac{3\gamma}{M-1}}\Bigg).\nonumber
%\end{align}
%The spectral efficiency is given by
%\begin{equation}
%	R=\frac{K\log_2M}{4} \quad \mathrm{bits/symbol},
%\end{equation}
%which is half of the efficiency of DCO-OFDM. 

\section{Numerical Results}\label{NumRes}
\begin{table}
	\renewcommand{\arraystretch}{1.1}
	\caption{The Parameter Setting}\vspace{-0.2cm}
	\label{table}
	\centering
	\resizebox{0.4\textwidth}{!}
	{\begin{tabular}{|c|c|c|}
			\hline
			Symbol & Definition & Value\\
			\hline\hline
			$\lambda_\mathrm{op}$ & Optical wavelength & $450$ nm\\
			\hline
			$\Upsilon_{\rm PDE}$ & The PDE of SPAD @ $\lambda_\mathrm{op}$ & $0.35$ \\ 
			\hline 
			$N_{\rm a}$ &  Number of SPAD pixels in the array & $8192$ \\
			\hline
			$\tau_{\rm d}$ & Dead time of SPAD &  $10$ ns\\
			\hline
			$P_{\rm B}$ & Background light power & $10$ nW\\
			\hline
			$\vartheta_{\rm DCR}$ & Dark count rate & $0.5$ MHz\\
			\hline
			$\varphi_{\rm AP}$ & Afterpulsing probability &  $0.75\%$\\
			\hline
			$\varphi_{\rm CT}$ & Crosstalk probability & $2.5\%$\\
			\hline 
			$P_{\rm max}$ & Maximal transmitted power & $20$ mW\\
			\hline 
			$K$ & Size of FFT and IFFT & $1024$ \\
			\hline
			$T_{\rm s}$ & Time-Domain sample duration  & $20$ ns\\
			\hline
	\end{tabular}}
\vspace{-0.3cm}
\end{table}
In this section, the numerical performance analysis of the SPAD-based OFDM system is presented. Unless otherwise mentioned, the parameters used in the simulation
are given in Table \ref{table}. The average transmitted optical power $\overline{P}_{\rm Tx}$ is given by (\ref{PTx}). By changing channel path loss $\zeta$,  various average optical power at the receiver $\overline{P}_{\rm Rx}=\zeta\,\overline{P}_{\rm Tx}$ can be achieved.

\begin{figure}[!t]
	\centering\includegraphics[width=0.45\textwidth]{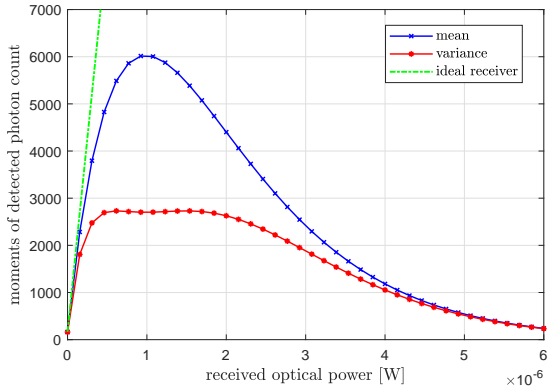}\vspace{-0.1cm}
	\caption{The mean and variance of the detected photon count of the considered SPAD receiver versus the received optical signal power. }\label{moment_SPAD_output}\vspace{-0.4cm}
\end{figure}
Fig. \ref{moment_SPAD_output} presents the mean and variance of the SPAD receiver output versus the received optical power, which are calculated based on (\ref{meanPQ}) and (\ref{varPQ}).  It is shown that due to the existence of the dead time, the average detected photon count has nonlinear relationship with the received power. 
%With the rise of the received power, the average detected photon count firstly increases and then decreases. 
%The maximal detected photon count of the SPAD receiver is around $6000$ in the considered system achieved when $\overline{P}_{\rm Rx}$ is around $1$ $\mu$W. 
The variance of the photon count is signal dependent and is less than the mean value. This is different from the ideal photon counting receiver, i.e., when $\tau_{\rm d}=0$, whose mean and variance of the detected photon count are identical and both increase linearly with the received optical power. 

%Therefore, the dead time can strongly reduce the detected photon count.            

%\begin{figure}[!t]
%	\centering\includegraphics[width=0.48\textwidth]{gain_factor.eps}
%	\caption{The gain factor $\alpha$ versus the received optical power with $\kappa=3$.} \label{gain_factor}
%\end{figure}
\begin{figure}[!t]
	\vspace{-0.09cm}
	\centering\includegraphics[width=0.45\textwidth]{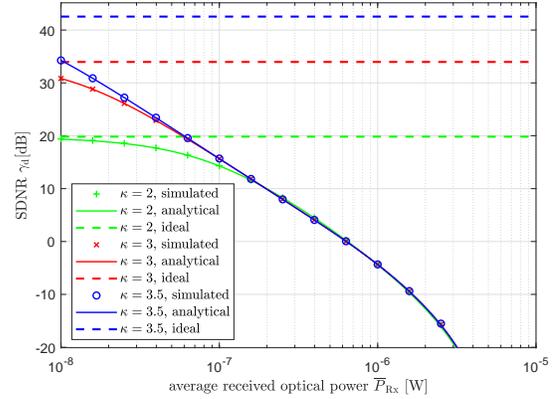}
	\caption{The SDNR versus the received optical power with various clipping level.}\vspace{-0.45cm} \label{SDNR}
\end{figure}
\begin{figure}[!t]
	\vspace{-0.1cm}
	\centering\includegraphics[width=0.43\textwidth]{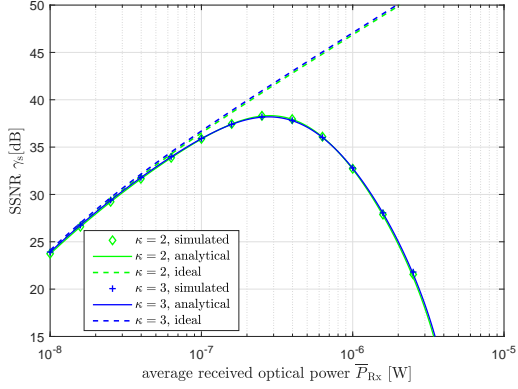}
	\caption{The SSNR versus the received optical power.} \vspace{-0.5cm}\label{SSNR}
\end{figure}

Fig. \ref{SDNR} shows the SDNR versus the received optical power with various clipping level. Note that for SPAD receiver, the considered nonlinear distortion contains both nonlinearities introduced by signal clipping at the transmitter and SPAD receiver as presented in (\ref{distortPQ}).  It is demonstrated that when the SPAD receiver is employed, with the increase of $\overline{P}_{\rm Rx}$, the SDNR decreases. This is because higher $\overline{P}_{\rm Rx}$ means wider utilized dynamic range of the receiver and hence severer receiver nonlinearity which leads to the lower SDNR. On the other hand, when the ideal photon counting receiver is used, the system is only influenced by signal clipping whose impacts do not change with $\overline{P}_{\rm Rx}$, resulting in the fixed SDNR over $\overline{P}_{\rm Rx}$. In addition, when $\overline{P}_{\rm Rx}$ is very small, the SDNR of the SPAD receiver converges to that of the ideal receiver due to the negligible SPAD nonlinearity. Fig. \ref{SDNR} further indicates that the system benefits from the higher $\kappa$, which can provide higher SDNR in low $\overline{P}_{\rm Rx}$ regime, because of the less nonlinearity induced by the signal clipping. However, when $\kappa$ is beyond $3$, since the nonlinear distortion caused by the signal clipping is sufficiently eliminated, the improvement introduced by high $\kappa$ reduces. In high $\overline{P}_{\rm Rx}$ regime, since the SDNR turns to be limited by the SPAD nonlinearity whose impact is similar when various $\kappa$ is employed,  the same SDNR is achieved regardless of the value of $\kappa$. Fig. \ref{SSNR} presents the SSNR versus the received optical power. It is shown that different from the SDNR, the SSNR is not sensitive to the varying $\kappa$. For SPAD receiver, with the rise of $\overline{P}_{\rm Rx}$, SSNR initially increases but then drops because of the severer SPAD nonlinearity. This is distinct from the ideal linear detector whose SSNR monotonically increases with the rise of $\overline{P}_{\rm Rx}$. It is also worth noting that in Fig. \ref{SDNR} and  Fig. \ref{SSNR} the analytical results of SDNR and SSNR perfectly match with the Monte Carlo simulation results, which validates our analytical derivations. 
\begin{figure}[!t]
	\centering\includegraphics[width=0.45\textwidth]{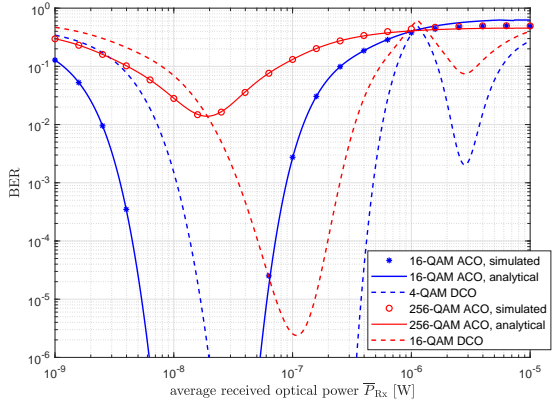}
	\caption{The BER versus the received optical power for SPAD-based OWC system with OFDM. }\vspace{-0.4cm} \label{ACO_DCO_OFDM_comp}
\end{figure}

%\begin{figure}[!t]
%	\centering\includegraphics[width=0.5\textwidth]{BER_result.eps}
%	\caption{The BER versus the received optical power when $16$-QAM is
%		employed.} \label{BER_result}
%\end{figure}

%\begin{figure}[!t]
%	\centering\includegraphics[width=0.5\textwidth]{BER_result_M32.eps}
%	\caption{The BER versus the received optical power when $32$-QAM is
%		employed.} \label{BER_result_M32}
%\end{figure}
The SNR of the considered SPAD-based OWC system with ACO-OFDM is given in (\ref{gamma}), based on which the BER result can be achieved. Fig. \ref{ACO_DCO_OFDM_comp} demonstrates the BER performance versus $\overline{P}_{\rm Rx}$ with $\kappa=3$. Two modulation schemes, i.e., $16$-QAM and $256$-QAM are considered as examples. The theoretical and simulation results again confirm a close match. It is illustrated that with the increase of $\overline{P}_{\rm Rx}$, BER firstly drops and then increases. This is because in small $\overline{P}_{\rm Rx}$ regime, the system is limited by the shot noise, the initial increase of $\overline{P}_{\rm Rx}$ can effectively increase SSNR  as shown in Fig. \ref{SSNR} and produce lower BER. However, in high $\overline{P}_{\rm Rx}$ regime, the gain of SSNR is overtaken by the SPAD nonlinear distortion. As presented in Fig. \ref{SDNR}, higher $\overline{P}_{\rm Rx}$ brings less SDNR, which results in higher BER. In addition, employing larger modulation order, e.g., $256$-QAM, introduces worse BER performance, as expected.  
Besides the system with ACO-OFDM, the performance of the system with DCO-OFDM is also considered for comparison. The detailed performance analysis of SPAD-based OWC system with DCO-OFDM is presented in \cite{arxivHuang}.  Because the spectral efficiency of the ACO-OFDM is half of the DCO-OFDM, to make a fair comparison between the schemes, the $M$-QAM ACO-OFDM should be compared with $\sqrt{M}$-QAM DCO-OFDM. It is illustrated that in the lower $\overline{P}_{\rm Rx}$ regime, $16$-QAM ACO-OFDM is more power efficient and requires around $4$ dB less optical power than $4$-QAM DCO-OFDM to achieve the same BER. However, in the high power regime, the latter in turn outperforms the former. 
This is because for the same $\overline{P}_{\rm Rx}$, ACO-OFDM signal spans over wider dynamic range compared to DCO-OFDM signal and thus experiences stronger receiver nonlinear distortion, which degrades its performance in high power scenarios. Note that for DCO-OFDM the extra dip of BER in high power scenario was explained in \cite{arxivHuang}. In addition, the comparison between $256$-QAM ACO-OFDM and $16$-QAM DCO-OFDM indicates that the superiority of ACO-OFDM in the low power regime drops when larger constellations are employed. Therefore, in the practical implementation, the employed OFDM schemes should be designed by considering both the received optical power and spectral efficiency requirement.     \vspace{-0.1cm}

\section{Conclusion}
ACO-OFDM can be used in SPAD-based OWC systems to achieve a good compromise between the high spectral efficiency and energy efficiency. In this work, a theoretical performance analysis of SPAD-based OWC systems with ACO-OFDM is presented. The analytical expressions of SNR and BER are derived which match with the Monte Carlo simulation results perfectly. 
%Therefore, the presented analytical results provide an effective and accurate way to estimate system performance of practical SPAD-based systems with ACO-OFDM. 
Through extensive numerical results, the impact of the SPAD nonlinearity on the system performance is investigated. It is further demonstrated that in the lower power regimes, ACO-OFDM is superior to DCO-OFDM (e.g., $4$ dB power gain achieved by $16$-QAM ACO-OFDM over $4$-QAM DCO-OFDM); whereas, in the high power regimes DCO-OFDM is more preferable.  
 \vspace{-0.1cm}
\bibliographystyle{IEEEtran}
\bibliography{IEEEabrv,SPAD_OFDM_Shenjie}
\end{document}